\documentclass[10pt]{article}
\usepackage{amsmath}
\usepackage{amssymb}

\usepackage{graphicx}

\usepackage{cite}
\usepackage{color}

\usepackage{setspace} 
\usepackage[labelfont=bf,labelsep=period,justification=raggedright]{caption}

\makeatletter
\renewcommand{\@biblabel}[1]{\quad#1.}
\makeatother

\date{}

\usepackage{amsfonts}
\usepackage[normalem]{ulem}
\usepackage[mathlines, displaymath, right]{lineno}
\usepackage{bm}

\begin{document}

% Title must be 150 characters or less
\begin{flushleft}
{\Large
\textbf{Interaction Mechanisms Quantified from Dynamical Features of Frog Choruses}
}
\\
Kaiichiro Ota$^{1, 2}$, 
Ikkyu Aihara$^{3, \ast}$, 
Toshio Aoyagi$^{2, 4}$

\bf{1} Cybozu, Inc., Tokyo, Japan\\
\bf{2} JST CREST, Tokyo, Japan\\
\bf{3} Graduate School of Systems and Information Engineering, University of Tsukuba, Tsukuba, Japan\\
\bf{4} Graduate School of Informatics, Kyoto University, Kyoto, Japan\\
$\ast$ E-mail: aihara@cs.tsukuba.ac.jp
\end{flushleft}

\clearpage

%\begin{keywords}
%Animal Behavior, Synchronization, Phase Oscillator Model, Bayesian Estimation
%\end{keywords}

\section*{Abstract}

Interaction mechanism in the acoustic communication of actual animals 
is investigated by combining mathematical modeling and empirical data.
Here we use a deterministic mathematical model (a phase oscillator model)
to describe the interaction mechanism underlying the choruses of male Japanese tree frogs ({\it Hyla japonica})
in which the male frogs attempt to avoid call overlaps with each other due to acoustic communication.
The mathematical model with a general interaction term is identified by a Bayesian approach 
from multiple audio recordings on the choruses of three male frogs.
The identified model qualitatively reproduces the stationary and dynamical features of the empirical data, 
supporting the validity of the model identification.
In addition, we quantify the magnitude of attention paid among the male frogs from the identified model, 
and then analyze the relationship between the attention and behavioral parameters by using a statistical model.
The analysis demonstrates the biologically valid relationship 
about the negative correlation between the attention and inter-frog distance, 
and also indicates the existence of a behavioral strategy that 
the male frogs selectively pay attention towards a less attractive male frog 
so as to utilize the advantage of their attractiveness for effective mate attraction.

\clearpage

\section{Introduction}

Animals show various types of behavior in the form of a swarm.
For instance, fish and birds construct a robust and flexible flock \cite{Vicsek_2012}.
To maintain  the flock, they need to synchronize their velocity and direction with each other. 
On the other hand, various animals (e.g., mammals, birds, anurans, and insects) 
aggregate in a certain area and utilize acoustic signals for communication and also mate identification 
\cite{Simmons_2002, Bird_2003, Gerhardt_2002, Wells_2007}.
Experimental studies demonstrate that these animals tend to alternate their acoustic signals with each other 
\cite{Takahashi_2013, Demartsev_2018, Gerhardt_2002, Wells_2007}.
Because such alternating behavior reduces the acoustic interference of their signals,
they are likely to effectively communicate or make themselves more conspicuous within a swarm.
Thus, synchronization and alternation are abundant in an animal swarm, 
and can work as an important indicator that determines the quality of their behavior.

To synchronize or alternate behavior, animals must recognize a specific target in a swarm.
Such selective attention is reported in various systems.
For example, humans pay attention to one of talking people in noisy environment like a party \cite{Bronkhorst_2000}; 
fish and birds attend their neighbor in a flock \cite{Fish_2011, Bird_2008};
bats pay attention to specific targets during a prey capture \cite{Fujioka_2016};
male insects and frogs call alternately with their neighbor \cite{Gerhardt_2002, Wells_2007}.
To understand the roles of the selective attention, it is essential for us 
to quantify their interaction mechanisms 
and compare it with the behavior of an individual animal.

\begin{figure*}
  \begin{center}
    %\begin{tabular}{cc}
	\includegraphics[width=1.0\textwidth]{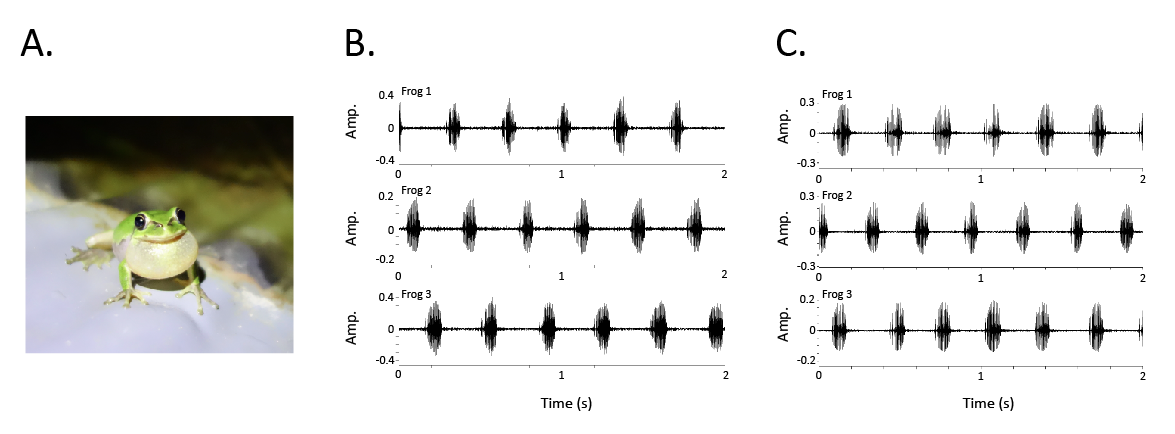}
	\end{center}
    %\end{tabular}
	\caption{
	Audio data on the choruses of three male Japanese tree frogs.
	(A) Photograph of a calling frog.
	(B) Tri-phase synchronization of three frogs.
	(C) Clustered anti-phase synchronization of three frogs.
	The male frogs tend to avoid call overlaps with each other.
	These figures are obtained from the empirical data of our previous study (Ref.[11]).
	}
	\label{fig:0}
\end{figure*}

This study aims to quantify interaction mechanisms inherent in the choruses of male Japanese tree frogs ({\it Hyla japonica}) 
by identifying a deterministic mathematical model (a phase oscillator model) from empirical data. 
Japanese tree frogs are observed widely in Japan, and breed mainly at a paddy field from April to July \cite{Maeda_1999}.
The male frogs form a lek, and produce successive calls to attract conspecific females.
Indoor experiments with multiple male Japanese tree frogs demonstrate
that they avoid call overlaps with each other in the forms of anti-phase synchronization of two frogs,
tri-phase synchronization of three frogs, 
and clustered anti-phase synchronization of three frogs
\cite{Aihara_2009, Aihara_2011} (see Figure \ref{fig:0} for examples of three frogs).
Moreover, field observations reveal two-cluster synchronization with a larger number of the male frogs 
in which each pair of neighbors tend to call alternately in their natural habitat \cite{Aihara_2014, Aihara_2016}.
Because the temporal overlap of acoustic signals generally mask the information included in each call,
these alternating behavior would be important for the male frogs to 
effectively advertise themselves towards a conspecific female 
\cite{Bee_2013, Aihara_2009, Aihara_2011, Aihara_2014, Aihara_2016}.
%Note that such an avoidance of call overlaps can be also observed 
%in insects \cite{Gerhardt_2002}, monkeys \cite{Takahashi_2013} and meercats \cite{Demartsev_2018}, 
%which demonstrates the importance of the alternating behavior 
%in the context of the acoustic communication of animals.

This paper is organized as follows:  
(1) a deterministic mathematical model (a phase oscillator model \cite{Kuramoto_1984}) is identified 
for each pair of male Japanese tree frogs from our empirical data according to a Bayesian approach, 
(2) the attention paid among the male frogs is quantified on the basis of the identified model,
and (3) the relationship between the quantified attention and behavioral parameters is analyzed by using a statistical model.

\section{Results}

To investigate interaction mechanisms inherent in the acoustic communication of actual animals, 
we analyze empirical data of male Japanese tree frogs that were obtained 
from our previous experiments and  data analysis \cite{Aihara_2011, RSOS_2018} 
(see Sec. \ref{sec:method_phase} for details). 
In each experiment, we randomly captured three male frogs at a field site, and then placed them along a straight line at the interval of $50$ cm in a stationary room. 
Spontaneous calling behavior of the male frogs was recorded by three microphones that were placed in the vicinity of each frog.
The audio data were analyzed by the method of independent component analysis,
and were separated into call signals of respective frogs.
Here we use four datasets of the audio data in which all the frogs stably called more than 1400 times 
in four hours, 
allowing us to precisely estimate a phase oscillator model by utilizing the large sample size of call timing.
%The experiment was carried out at 44 times in a stationary room between 2008 and 2009.
%Then, we carefully checked the audio data of all the experiments, and confirmed that three frogs stably called in four experiments.

\begin{figure*}
  \begin{center}
    %\begin{tabular}{cc}
	\includegraphics[width=0.9\textwidth]{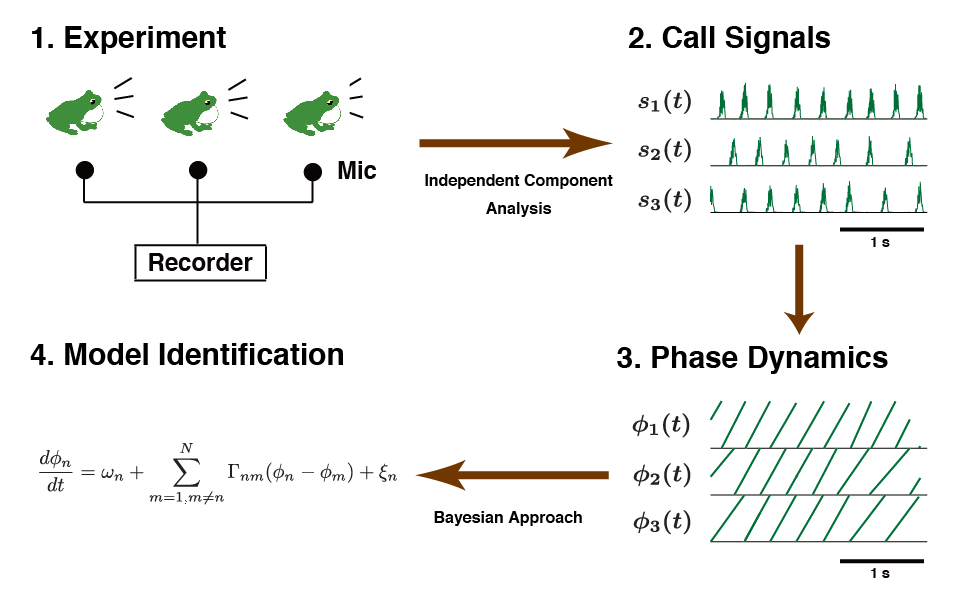}
	\end{center}
    %\end{tabular}
	\caption{
	Schematic diagram on the identification of a phase oscillator model.
	In this study, we utilize the audio data of male Japanese tree frogs obtained 
	from our previous study \cite{Aihara_2011}.
	%The audio data were separated into the call signals of respective frogs 
	%by the method of independent component analysis.
	Phase dynamics is estimated from the audio data.
	We then identify a phase oscillator model by analyzing the phase dynamics 
	according to a Bayesian approach, 
	which allows us to infer the interaction mechanisms among the actual frogs. 
	}
	\label{fig:1}
\end{figure*}

\begin{figure*}
  \begin{center}
    %\begin{tabular}{cc}
	\includegraphics[width=0.9\textwidth]{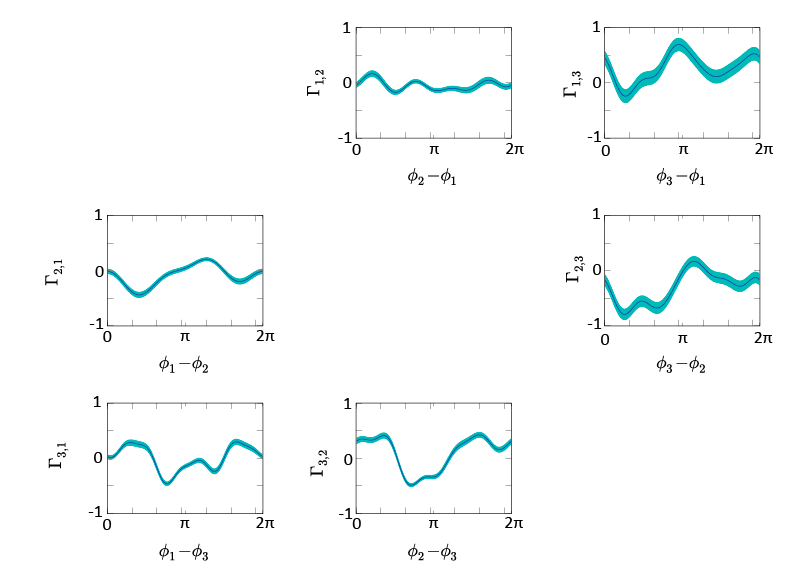}
	\end{center}
    %\end{tabular}
	\caption{
	Unidirectional interaction terms of a phase oscillator model that are identified from the empirical data 
	by a Bayesian approach.
	In this study, the interaction term $\Gamma_{n, m}$ describes how the $n$th frog controls its call timing 
	by responding to the calls of the $m$th frog.
	Cyan region represents the $95\%$ confidence interval of the identified interaction term. 
	}
	\label{fig:2}
\end{figure*}

Figure \ref{fig:1} explains how we identify a phase oscillator model from the empirical data.
%In this study, we identify a phase oscillator model from the empirical data according to the flowchart of Figure \ref{fig:1}.
A phase oscillator model is a well-known deterministic mathematical model 
derived from simple assumptions about periodicity and interaction \cite{Kuramoto_1984} 
(see Sec.\ref{sec:phase_model} for the details of a phase oscillator model)
that are essential for the acoustic communication of animals.
Various experimental and theoretical studies have shown that 
a phase oscillator model can qualitatively reproduce the synchronization phenomena in actual systems 
\cite{Strogatz, Nenkin, BZ, Walk} including alternating chorus patterns of male Japanese tree frogs \cite{Aihara_2009, Aihara_2011, Aihara_2014, RSOS_2018}.
%\MOD{(Explanation of the merits for the use of a phase oscillator model. The validity to use the model for the frogs.)}
In this study, we first calculate a phase $\phi_{n}(t_{i})$ ($n=1$, $2$, $3$) with discrete time $t_{i}$ for the $n$th frog 
from the separated audio data according to Eq.~(\ref{eq:definition_phase}).
Then, a Bayesian method \cite{Ota_arXiv} is applied to the time series data of $\phi_{n}(t_{i})$ 
so as to estimate the parameters of a phase oscillator model 
(i.e., an interaction term $\Gamma_{nm}(\phi_{n}(t_{i})-\phi_{m}(t_{i}))$, a natural frequency $\omega_{n}$, 
and a noise intensity $\sigma_{n}$). %in Eq.~(\ref{eq:phase_oscillator_model})).
Figure \ref{fig:2} shows a representative result of the model identification that is obtained 
from a single dataset with three frogs.
Green bold lines represent the interaction terms with the $95\%$ confidence interval. 
%which demonstrates that we succeed in estimating these terms with a high precision.
%Note that the interaction term $\Gamma_{nm}(\phi_{n}(t)-\phi_{m}(t))$ describes how the $n$th frog responds to calls of the $m$th frog.

\begin{figure*}
  \begin{center}
    %\begin{tabular}{cc}
	\includegraphics[width=0.9\textwidth]{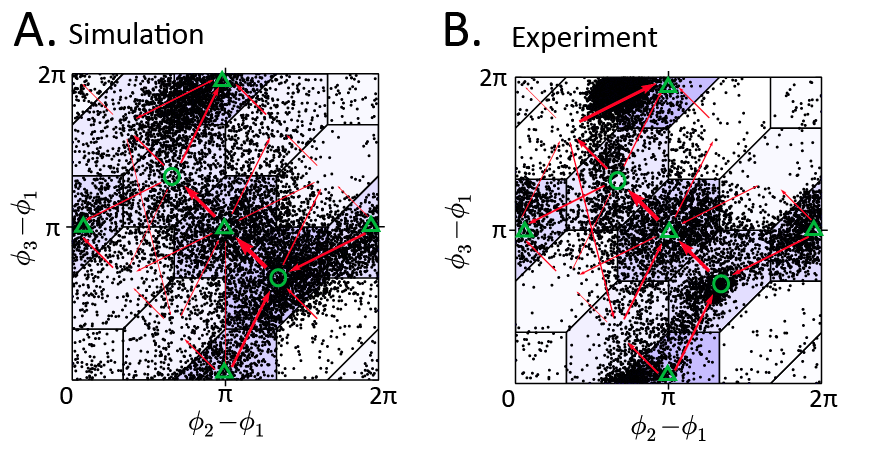}
	\end{center}
    %\end{tabular}
	\caption{
	Phase differences between the calls of three male frogs that are obtained from 
	(A) numerical simulation of the identified model and (B) behavioral experiment of actual frogs.
	%For the simulation, we use the phase-oscillator models identified from the behavioral experiment.
	Each black dot represents a set of the phase differences $\phi_{2}-\phi_{1}$ and $\phi_{3}-\phi_{1}$. 
	Circle and triangle depict the regions of tri-phase synchronization 
	and clustered anti-phase synchronization, respectively.
	Red arrows represent the transitions among the synchronization states. 
	The width of each arrow is proportional to the number of the transitions. 
	}
	\label{fig:3}
\end{figure*}

\begin{figure*}
  \begin{center}
    %\begin{tabular}{cc}
	\includegraphics[width=0.9\textwidth]{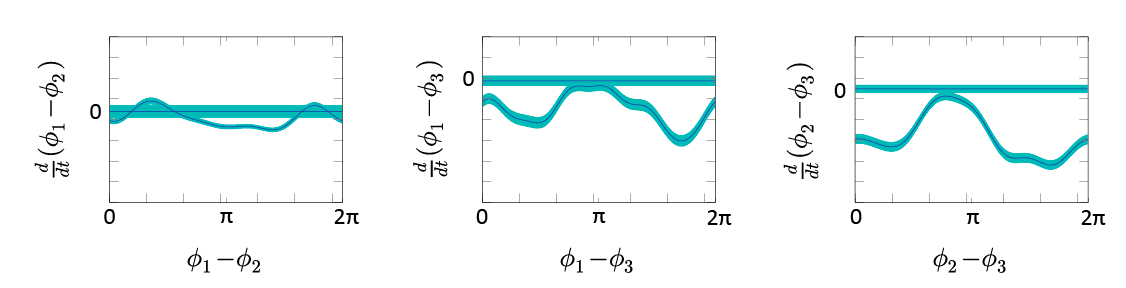}
	\end{center}
    %\end{tabular}
	\caption{
	Critical states of the identified model.
	Cyan region represents the $95\%$ confidence interval of $d(\phi_{n}-\phi_{m})/dt$ 
	that is estimated from the empirical data by a Bayesian approach.
	It is demonstrated that $d(\phi_{1}-\phi_{3})/dt$ and $d(\phi_{2}-\phi_{3})/dt$ have critical states
	while $d(\phi_{1}-\phi_{2})/dt$ has equilibrium states.
	}
	\label{fig:4}
\end{figure*}

To confirm the validity of the model identification, 
we perform numerical simulation by using the identified model and compare it with empirical data.
Note that we set only one parameter $\sigma_{n}$ to be slightly larger than its estimated value throughout the following analysis  
because this parameter was very likely to be estimated at a smaller value 
because of the intermittency of the chorus over a long time scale (see Sec.~\ref{sec:method_phase} for details) 
that is out of the scope of the phase oscillator model.
Figure \ref{fig:3}A represents the scatter plot of phase differences 
that are obtained from numerical simulation of the identified model.
By contrast, Figure \ref{fig:3}B shows the scatter plot of phase differences 
that are directly calculated from our empirical data. %according to Eq.(\ref{eq:definition_phase}).
Here we plot the phase differences only when one of the phases hits $0$, 
which is consistent with our method for calculating a phase difference from discrete call timing \cite{Aihara_2011, Aihara_2016}.
The comparison of Fig. \ref{fig:3}A and B demonstrates that the identified model can qualitatively reproduce the experimental result, 
supporting the validity of our model identification.
In addition, our empirical data of Fig.~\ref{fig:3}B shows a complex transition 
among clustered anti-phase synchronization and tri-phase synchronization, 
which is consistent with our previous study \cite{Aihara_2011}.
To  investigate the mechanism of such a transition, 
we further analyze the relationship between a phase difference $\phi_{n}-\phi_{m}$ 
and its differential $d(\phi_{n}-\phi_{m})/dt$ by using the identified model (see Sec.\ref{sec:critical_state}).
Figure~\ref{fig:4} demonstrates that $d(\phi_{1} - \phi_{3})/dt$ and $d(\phi_{2} - \phi_{3})/dt$ have critical states  
while $d(\phi_{1} - \phi_{2})/dt$ has equilibrium states.
Because the transitions among multiple states can be driven by added noise under the existence of critical states 
in various dynamical systems, 
this property about critical states is likely to be the origin of the transitions 
observed in the empirical data of the frog choruses.
% we may need some references here.

\begin{figure*}
  \begin{center}
    %\begin{tabular}{cc}
	\includegraphics[width=0.9\textwidth]{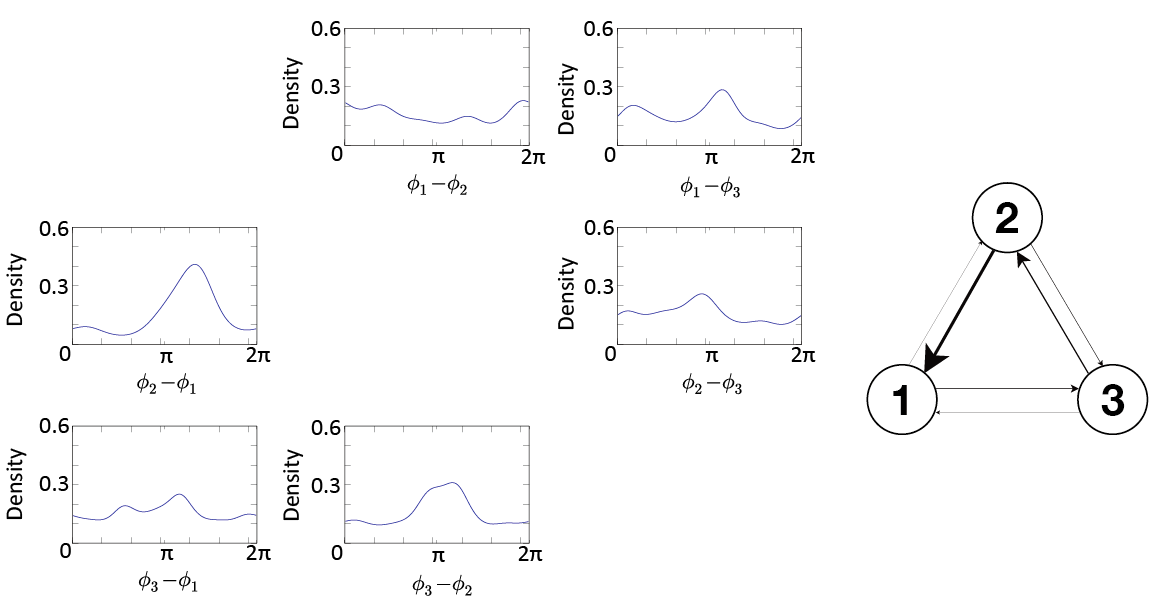}
	\end{center}
    %\end{tabular}
	\caption{
	Selective attention quantified from the identified model.
	(Left) Stationary distribution of the phase differences 
	that is obtained from the Fokker-Plank equation of the identified model.
	(Right) Schematic diagram of selective attention 
	that is quantified by using the Kullback-Leibler divergence 
	of the stationary distribution from uniform distribution.
	Line width represents the magnitude of attention paid among the male frogs.
	}
	\label{fig:5}
\end{figure*}

Next, we quantify the magnitude of the attention paid among the male frogs based on the result of the model identification.
The stationary distribution of each phase difference is first calculated by numerically solving the Fokker-Planck equation
of the identified model (see Sec.~\ref{sec:FP_KL}).
Figure \ref{fig:5} demonstrates that some distributions have a sharp peak.
For instance, the distribution of $\phi_{2} - \phi_{1}$ has a sharp peak around $\pi$, 
meaning that the $2$nd frog attempted to call alternately with the $1$st frog.
Then, we calculate the Kullback-Leibler divergence of the calculated distribution 
from uniform distribution for each pair of the male frogs 
(see Sec.~\ref{sec:FP_KL}), 
which corresponds to the quantification of the attention.
The right panel of Fig.~\ref{fig:5} shows the result of the quantification 
in which line width represents the magnitude of the attention. 
Consequently, an asymmetric structure is observed in the attention paid among the male frogs.
For example, the $1$st frog pays strong attention to the $2$nd frog 
while the $2$nd frog pays just weak attention to the $1$st frog.

To further examine the validity of the model identification, 
we analyze the relationship between the attention and behavioral parameters 
by using a generalized linear mixed model (see Sec.~\ref{sec:statics}). %\cite{Faraway_2006}
In this analysis, the magnitude of the attention is treated as a response variable,
and three behavioral parameters (i.e., an inter-frog distance, the probability for taking a chorus leader (leader probability), and an inter-call interval)
are treated as explanatory variables of fixed factors. 
These variables are calculated from four datasets with $12$ frogs 
in which all the frogs stably produced calls. 
%(see Sec. \ref{sec:method_phase} and Supplementary Information).
% In addition, Frog ID and experimental date are treated as explanatory variables of random factors. 
Figure~\ref{table:1} shows the posterior mean and the 95$\%$ confidence interval of the coefficients of respective fixed factors. %(see Sec.~\ref{sec:statics} for detail).
This analysis demonstrates that 
(1) the inter-frog distance has a negative effect on the attention, 
meaning that male frogs pay more attention to their neighbor,
and also indicates that (2) the leader probability and the inter-call interval have negative and positive effects on the attention, respectively.

\section{Discussions}

In this study, we quantify interaction mechanisms in the chorus of male Japanese tree frogs 
by identifying a phase oscillator model from empirical data.
The identified model qualitatively reproduces the stationary and dynamical features of the frog choruses, 
which supports the validity of our model identification.
Then, the magnitude of the attention paid among the male frogs is quantified on the basis of the identified model.
%The analysis using a statistical model indicates that the magnitude of the attention is correlated 
%with the behavioral parameters of the male frogs.
To our knowledge, this is the first study that shows the evidence of selective attention inherent in an animal chorus  
by combining empirical data with a deterministic mathematical model.

\begin{figure*}
  \begin{center}
    %\begin{tabular}{cc}
	\includegraphics[width=1.0\textwidth]{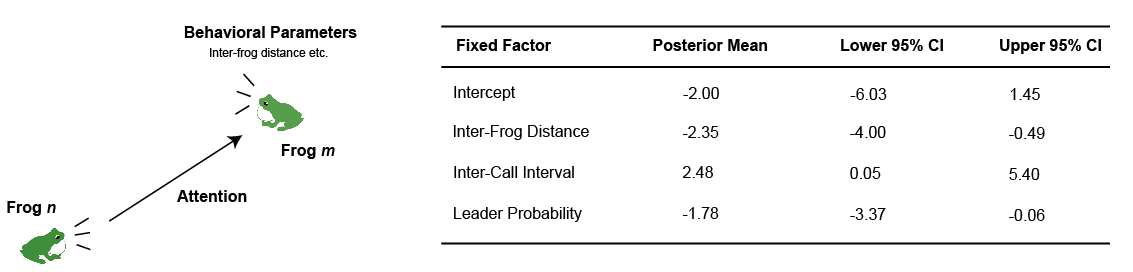}
	\end{center}
    %\end{tabular}
	\caption{
	Relationship between selective attention and behavioral parameters examined by a statistical model (GLMM).
	The magnitude of attention is treated as a response variable; 
	three behavioral parameters 
	(i.e., an inter-frog distance, an inter-call interval and leader probability) 
	are treated as explanatory variables of fixed factors.
	This result is obtained from the empirical data that consist of four datasets with $12$ frogs in total. 
	%(the empirical data of call timing is available for all the datasets from Supplementary Information of this manuscript).
	}
	\label{table:1}
\end{figure*}

The relationship between the attention and the behavioral parameters (Figure~\ref{table:1})
gives perspectives on the choruses of male Japanese tree frogs.
\begin{itemize}
\item {\bf Inter-frog distance:} 
Our analysis using a statistical model demonstrates the negative relationship between the attention and the inter-frog distance, 
which means that male Japanese tree frogs pay more attention to their neighbor.
This is consistent with previous studies reporting that a neighboring pair of males alternate their calls 
in various species of frogs and insects \cite{Gerhardt_2002, Wells_2007, Aihara_2016, Brush_1989, Jones_2014}.
Because sounds attenuate depending on distance, the calls of a neighboring pair should arrive at a higher intensity than the calls of a distant pair.
We speculate that such a sound attenuation depending on distance 
is the origin of the negative relationship between the attention and the inter-frog distance. 
\item {\bf Inter-call interval and leader probability:}
In various species of frogs and insects, females prefer a conspecific male 
that leads their chorus \cite{Grafe_1996, Snedden_1998} and produces calls at a higher repetition rate \cite{CallRate_1992}.
Our analysis indicates that male Japanese tree frogs pay more attention to a male that rarely leads their chorus and produces calls at a lower repetition rate.
These results suggest that the male frogs pay more attention to a less attractive male and then call alternately with him. 
Given that alternating chorus patterns can reduce the acoustic interference of their calls \cite{Bee_2013, Aihara_2009}, 
this feature would be important for male frogs to effectively advertise themselves towards conspecific females
by utilizing the advantage of their attractiveness over a neighboring male.
\end{itemize}
Thus, our analysis is likely to show the important features in frog choruses, 
which is valid in the contexts of the acoustic communication of male frogs as well as the strategy for mating.
However, it should be noted that 
the 95$\%$ confidence intervals of the coefficients of the inter-call interval and leader probability do not include but very close to $0$
while the 95$\%$ confidence interval of the coefficient of the inter-frog distance is obviously beyond $0$ (Figure~\ref{table:1}). 
This suggests that the effects of the leader probability and inter-call interval on the attention are marginal compared to the effect of the inter-frog distance.

The present methodology is widely applicable to the analysis on various types of communication in animals because of the following reasons: 
(1) a phase oscillator model is derived from the simple assumptions about periodicity and interaction 
both of which are ubiquitous in the communication of animals relying on various signals 
(e.g., sounds \cite{Gerhardt_2002, Wells_2007}, lights \cite{Buck_1968, Strogatz_2004}, visual display \cite{Backwell_1998, Backwell_1999}, and electric fields \cite{Markham_2009}), 
(2) our methodology only requires the timing of signal emissions so as to identify the mathematical model, 
(3) our methodology allows us to separately identify the unidirectional interaction term of a phase oscillator model,
and (4) the identified model qualitatively reproduces the stationary and dynamical features of empirical data.
An important point is that the interaction term of the phase oscillator model varies 
depending on the value of the phase difference that corresponds to the change in the inter-signal interval among individual animals.
This study demonstrates that such a dynamical property of the phase oscillator model can precisely capture not only the stationary distribution of the phase difference 
but also the dynamical feature of the transition among multiple synchronization states. 
We believe that this property of a dynamical model is advantageous compared to traditional methods (e.g., the calculation of the histogram of the inter-signal interval) 
when studying selective attention in animal communication that shows complicated deterministic dynamics.

%In particular, recent progress of sound processing techniques using a microphone-array system allows us 
%to reveal calling signals of animals such as frogs \cite{Grafe_1997, Simmons_2008, Jones_2014},
%birds \cite{Suzuki_2016}, bats \cite{Fujioka_2014} and dolphins \cite{Au_2003}; 
%the combination of these techniques with our methodology is promising for the .

\section{Materials and Methods}

\subsection{Estimation of phase dynamics from empirical data}
\label{sec:method_phase}

In this study, we use the empirical data of male Japanese tree frogs that were obtained from our previous experiments \cite{Aihara_2011}.
In each experiment, we randomly captured three individuals of the male frogs that were calling in a paddy field of Kyoto University (35°01' 57.2"N, 135°47'00.0"E).
Then, we put each frog in a small mesh cage, and placed three cages along a straight line at the interval of $50$cm in a stationary room.
After sunset, the spontaneous calling behavior of the male frogs was recorded in the stationary room by three microphones that were placed in the vicinity of each cage.
After the recording, the male frogs were released at the paddy field where they were captured.
The experiment was carried out $44$ times in total between 2008 and 2009.
In most trials of the experiments, some of three frogs rarely or did not call at all. 
Because this study aims to estimate the parameters of a mathematical model due to a statistical approach
and the precision of the parameter estimation generally depends on a sample size of the empirical data, 
we choose four datasets in which all the three frogs stably produced calls more than 1400 times in four hours;
the four datasets were obtained from the experiments performed on $26$th May, $16$th June, $17$th June in 2008, and $29$th May in 2009, respectively.
%in the room temperature of \MOD{temp1, temp2, temp3, and temp4}, respectively.
%The call timing of individual frogs obtained from the four datasets are available as the Supplementary Information of this manuscript.
All the experiments and collection of Japanese tree frogs were carried out within the facility of Kyoto University 
in accordance with the guideline approved by the Animal Experimental Committee of Kyoto University.
Then, the audio data of the four datasets were analyzed by the method of independent component analysis, 
and were separated into call signals of respective frogs \cite{Aihara_2011}.
In addition, we carefully checked the quality of the sound-source separation 
and excluded just three choruses from 143 choruses in which the separation did not work well
\cite{RSOS_2018}.
From the separated audio data, we estimated the call timing of respective frogs 
according to the method of Ref.\cite{Aihara_2011}.
%(All the data of the call timing is available from Supplementary Information of this manurscript).

For each dataset, the call timing of three male frogs is described as $T_{n, k}$ that represents the timing of the $k$th call vocalized by the $n$th frog ($n=1$, $2$ or $3$).
By using the call timing $T_{n, k}$, we calculate a phase $\phi_{n}(t_{i})$ for the $n$th frog at discrete time $t_{i}$ as follows \cite{Piko, Aihara_2011, Aihara_2014}:
\begin{eqnarray}
\phi_{n}(t_{i}) = 2\pi \frac{t_{i}-T_{n, k}}{T_{n, k+1}-T_{n, k}}.
\label{eq:definition_phase}
\end{eqnarray}
This calculation is performed for the set of $T_{n, k}$ and $T_{n, k+1}$ 
that satisfy the conditions of $T_{n, k} \le t_{i} < T_{n, k+1}$ and $T_{n, k+1}-T_{n, k} < 0.9$ sec. 
The first condition (i.e., $T_{n, k} \le t_{i} < T_{n, k+1}$) is necessary 
for restricting the phase in the range of $0 \le \phi_{n}(t_{i}) < 2\pi$.
The second condition (i.e., $T_{n, k+1}-T_{n, k} < 0.9$ sec) is assumed 
because of the property of the frog chorus 
in which male Japanese tree frogs intermittently start and stop their periodic calling behavior over a long time scale.
Namely, each frog periodically calls almost at the interval of $0.3$ sec for several tens of seconds, stays silent for several minutes, and then repeats this cycle \cite{RSOS_2018}.
Because such an intermittency over a long time scale is out of the scope of the phase oscillator model, we assume the second condition 
and then calculate the phase $\phi_{n}(t_{i})$ only during the periodic calling behavior. 
Consequently, the phase $\phi_{n}(t_{i})$ increases from $0$ to $2\pi$, 
and then reset to be $0$ when the $n$th frog produces a call during the periodic calling behavior. 
%Note that the threshold of $0.9$ sec is set to be sufficiently larger than the mean value of the inter-call interval in male Japanese tree frogs 
%so as to robustly obtain the time series data of the phase $\phi_{n}(t)$ from our empirical data 
%in which male frogs dynamically change their call timing by responding to the calls of other frogs.

\subsection{Phase oscillator model}
\label{sec:phase_model}

To reproduce the stationary and dynamical features in the choruses of male Japanese tree frogs, 
we use a phase oscillator model \cite{Kuramoto_1984} with additive noise as follows:
\begin{eqnarray}
\frac{d\phi_{n}(t)}{dt} &=& \omega_{n} + \sum_{m=1, m \neq n}^{N}\Gamma_{nm}(\phi_{n}(t)-\phi_{m}(t)) + \xi_{n}(t),
\label{eq:phase_oscillator_model}
\end{eqnarray}
where $\phi_{n}(t) \in [0, 2\pi)$ ($n = 1$, $2$, ..., $N$) is the phase of the $n$th frog. 
$\omega_{n}$  is a positive parameter that describes the intrinsic angular velocity of the $n$th frog.
We then assume that the $n$th frog produces a call at $\phi_{n}(t) = 0$, which is consistent with the definition of Eq.~(\ref{eq:definition_phase}).
Consequently, $2\pi/\omega_{n}$ gives the intrinsic inter-call interval of the $n$th frog
\cite{Aihara_2009, Aihara_2011, Aihara_2014, RSOS_2018}. 
$\Gamma_{nm}(\phi_{n}(t)-\phi_{m}(t))$ is the unidirectional interaction term of the phase oscillator model, 
and is defined as a $2\pi$-periodic function of $\phi_{n}(t)-\phi_{m}(t)$ \cite{Kuramoto_1984}.
In the context of acoustic communication amomg male frogs, this term represents how the $n$th frog controls his call timing by responding to the calls of the $m$th frog 
\cite{Aihara_2009, Aihara_2011, Aihara_2014}.
$\xi_{n}(t)$ is the term of additive noise.
We assume that this term is given by independent white Gaussian noise with magnitude $\sigma_{n}$ 
that satisfies the relationship $<\xi_{n}(t)\xi_{n}(s)>=\sigma_{n} \delta(t-s)$.
Note that we utilize this term $\xi_{n}(t)$ so as to better fit the model parameters from empirical data that contain noisy component. 
%that were obtained from our experiments using actual frogs (see Sec.~\ref{sec:method_phase}). 

\subsection{Identification of a phase oscillator model}
\label{sec:model-identification}

Given the time series data of the phase $\phi_{n}(t_i)$ with discrete time $t_i$ 
that was obtained from our experiments using actual frogs (see Sec.~\ref{sec:method_phase}),
we estimate the unknown parameters of the phase oscillator model according to the Bayesian method of Ref.~\cite{Ota_arXiv}.
In the method, we first define the following likelihood function from the time series data of the phase $\phi_{n}(t_i)$:
\begin{eqnarray}
%\begin{align*}
    L_n = \prod_i \mathcal{N} \Big[ \frac{\phi_n(t_{i+1}) - \phi_n(t_i)}{t_{i+1} - t_i} \Big| \omega_n 
    + \sum_{m=1,m \neq n}^3 \Gamma_{nm}(\psi_{nm}(t_i)), \frac{\sigma_n^2}{t_{i+1} - t_i} \Big],
    \label{eq:likelihood}
\end{eqnarray}
%\end{align*}
%%
where $\mathcal{N}(x|m, s^2)$ represents a Gaussian function with mean $m$ and variance $s^2$, 
and $\psi_{nm}$ denotes the phase difference $\phi_n - \phi_m$.
Because each interaction term is defined as a $2\pi$-periodic function of the phase difference $\psi_{nm}$ \cite{Kuramoto_1984},
it can be expanded into a Fourier series as follows:
\begin{equation}
\Gamma_{nm} (\psi_{nm}) = \sum_{k=1}^{M} \big(a_{nm}^{(k)} \cos k\psi_{nm} + b_{nm}^{(k)} \sin k\psi_{nm} \big).
\label{eq:interaction_fourier}
\end{equation}
%\begin{equation}
%    \Gamma_{nm}(\psi_{nm}) = a_{nm}^{(1)} \cos \psi_{nm} + b_{nm}^{(1)} \sin \psi_{nm}
%    + a_{nm}^{(2)} \cos 2\psi_{nm} + b_{nm}^{(2)} \cos 2\psi_{nm} \\
%    + \cdots + a_{nm}^{(M)} \cos M\psi_{nm} + b_{nm}^{(M)} \sin M \psi_{nm}.
%    \label{eq:interaction_fourier}
%\end{equation}
%%
We apply a standard model comparison method to determine the maximum order $M$ in Eq.(\ref{eq:interaction_fourier}).
That is, we choose $M$ such that a marginal likelihood function $\mathcal{L}_M$ is maximized.
Technically, we computed and compared $\mathcal{L}_M$ for $M = 1, 2, \dots, 30$.
For our data, the maximum order $M$ was chosen between 1 and 9.

Then, the parameters to be estimated are
$\omega_n, a_{nm}^{(1)}, \dots, a_{nm}^{(M)}, b_{nm}^{(1)}, \dots, b_{nm}^{(M)}$ and $\sigma_n$,
which are denoted by a shorthand notation $\bm{c}_n$.
To estimate $\bm{c}_n$ in a Bayesian framework, 
we use a reasonable conjugate prior distribution $p_\mathrm{prior}(\bm{c}_n)$,
which is a Gaussian-inverse-gamma function.
Bayes' theorem then gives the posterior parameter distribution 
%which can be understood as the probability distribution of the unknown parameters 
%that is estimated from our empirical data, 
as follows:
\begin{equation}
    p_\mathrm{post}(\bm{c}_n) \propto L_n(\bm{c}_i) p_\mathrm{prior}(\bm{c}_n).
    \label{eq:posterior}
\end{equation}
Because the prior distribution $p_\mathrm{prior}(\bm{c}_n)$ is conjugate to the likelihood $L_n(\bm{c}_i)$,
we can easily calculate the posterior distribution $p_\mathrm{post}(\bm{c}_n)$ 
%according to the method of Ref.\cite{Ota_arXiv}.
(see Ref.\cite{Ota_arXiv} and its supplemental material for detail).

\subsection{Examination of critical states}
\label{sec:critical_state}

Empirical data on the choruses of male Japanese tree frogs demonstrate the complicated transition among multiple synchronization states 
(Fig.\ref{fig:3}B) \cite{Aihara_2011}.
Here we examine the origin of this transition on the basis of the estimated model.
Equation (\ref{eq:phase_oscillator_model}) without a noise term 
yields the time evolution of the phase difference $\psi_{nm} \equiv \phi_n - \phi_m$ as follows:
\begin{equation}
    G_{nm}(\psi_{nm}) \equiv \frac{d(\phi_n - \phi_m)}{dt} = \omega_n - \omega_m + \Gamma_{nm}(\psi_{nm}) - \Gamma_{mn}(-\psi_{nm}),
    \label{eq:phase_difference}
\end{equation}
where $\Gamma_{nm}(\psi_{nm})$ is the maximum a posteriori (MAP) estimation of the interaction term
%(in other words, the most probable function of the interaction term) 
that was obtained from the analysis of Sec.~\ref{sec:model-identification}.
%Note that we exclude the effect of the noise term $\xi_{n}(t)$ in this analysis
%so as to first focus on the deterministic aspect of the estimated model.
This function $G_{nm}(\psi_{nm})$ quantifies how the phase difference changes in time 
under the mutual interaction between the $n$th and $m$th frogs without the effect of another frog.
If $G_{nm}(\psi_{nm})$ has a zero-crossing (a stable equilibrium) at $\psi_{nm} = \psi_{nm}^*$ , 
the phase difference approaches $\psi_{nm}^*$ and then stays near the value forever.
Such stationary dynamics typically come from strong interaction.
By contrast, if $G_{nm}(\psi_{nm})$ has no zero-crossing but is very close to zero at a certain point $\psi_{nm} = \psi_{nm}^*$  (a critical state), 
the phase difference stays near $\psi_{nm}^*$ for a certain amount of time and eventually departs the point.
Such critical dynamics typically come from moderate interaction.

\subsection{Quantification of selective attention}
\label{sec:FP_KL}

To assess selective attention among male frogs, we evaluate the stochastic feature of the estimated model 
by taking the effect of noise into consideration.
%($+$ add explanation on the difference between the attention and the interaction term).
For example, to evaluate the attention from the $1$st frog to the $2$nd frog,
we analyze the following model only with the unidirectional interaction term $\Gamma_{12}(\phi_1(t) - \phi_2(t))$:
\begin{eqnarray}
	\frac{d\phi_1}{dt} = \omega_1 + \Gamma_{12}(\phi_1(t) - \phi_2(t)) + \xi_1(t),
	\label{eq:frog1_to_2}\\
	\frac{d\phi_2}{dt} = \omega_2 + \xi_2(t).
	\label{eq:frog2_to_1}
\end{eqnarray}
These equations describe the situation that the $1$st frog exposed to the noise term $\xi_1(t)$ 
pays attention to the $2$nd frog according to the interaction term $\Gamma_{12}(\phi_1(t) - \phi_2(t))$ (Eq.~(\ref{eq:frog1_to_2}))
but the $2$nd frog exposed to the noise term $\xi_2(t)$ does not pay any attention to the $1$st frog (Eq.~(\ref{eq:frog2_to_1})).
Hence, this is a concise mathematical model capturing the attention from the $1$st frog to the $2$nd frog 
under the effect of the noise $\xi_1(t)$ and $\xi_2(t)$.
Subtracting Eq.~(\ref{eq:frog2_to_1}) from Eq.~(\ref{eq:frog1_to_2}) yields the time evolution of the phase difference $\psi_{12} \equiv \phi_1(t) - \phi_2(t)$ as follows:
%The dynamics of the phase difference $\psi_{12} \equiv \phi_1(t) - \phi_2(t)$ is then written as the following stochastic differential equation from Eqs.~(\ref{eq:frog1_to_2}) and  (\ref{eq:frog2_to_1}) 
%%
\begin{equation}
    \frac{d\psi_{12}}{dt} = \omega_1 - \omega_2 + \Gamma_{12}(\psi_{12}) + \xi_1(t) + \xi_2(t).
    \label{eq:selective_attention}
\end{equation}
Then, we calculate the stationary distribution of the phase difference $\psi_{12}$
by numerically solving the Fokker-Planck equation of Eq.(\ref{eq:selective_attention})
that gives the time evolution of the distribution of the phase difference (i.e., $f(\psi_{12}, t)$),
\begin{equation}
    \frac{\partial f(\psi_{12}, t)}{\partial t}
    = -\frac{d\Gamma_{12}}{d\psi_{12}}(\psi_{12})f
    - [\omega_1 - \omega_2 + \Gamma_{12}(\psi_{12})]\frac{\partial f}{\partial \psi_{12}}
    + \frac{\sigma_1^2 + \sigma_2^2}{2}\frac{\partial^2 f}{\partial \psi_{12}^2},
    \label{eq:fokker_planck}
\end{equation}
until it converges. 
Subsequently, we can calculate the stationary distribution of the phase difference for all the pairs of male frogs, 
and then describe it as $\hat{f}(\psi_{nm})$.

The stationary distribution of a phase difference ($\hat{f}(\psi_{nm})$) allows us to quantify the degree of attention.
If there is no attention from the $n$th frog to the $m$th frog, the distribution of $\psi_{nm}$ is almost uniform; 
on the other hand, if the $n$th frog calls synchronously with the $m$th frog as the result of paying strong attention to him, 
%if a male frog pays strong attention to another frog and call synchronously, 
the distribution has a sharp peak. 
%the phase difference of their calls should lock to a certain value, which results in a sharply peaked stationary distribution.
In this study, we quantify such a sharpness of the distribution $\hat{f}(\psi_{nm})$ by using the Kullback-Leibler divergence 
from uniform distribution $u(\psi_{nm}) \equiv \frac{1}{2\pi}$ as follows:
\begin{equation}
    D_\mathrm{KL}(\hat{f}||u) = \int_0^{2\pi} \hat{f}(\psi_{nm})\log\frac{\hat{f}(\psi_{nm})}{u(\psi_{nm})} d\psi_{nm}.
    \label{eq:kullback_leibler}
\end{equation}
Consequently, a nearly uniform distribution is represented by $D_\mathrm{KL}(\hat{f}||u) \sim 0$
while a sharply peaked distribution is represented by $D_\mathrm{KL}(\hat{f}||u)$ that is much larger than $0$.
%Thus, the Kullback-Leibler divergence $D_\mathrm{KL}(\hat{f}||u)$ corresponds to the magnitude of selective attention in the frog choruses.

\subsection{Relationship between selective attention and behavioral parameters}
\label{sec:statics}

To further examine the validity of the model identification, 
we analyze the relationship between selective attention and behavioral parameters.
Here we focus on the following behavioral parameters of male Japanese tree frogs:
(1) an inter-frog distance, (2) an inter-call interval, and (3) leader-follower relationship.
An inter-frog distance represents the distance between each pair of male frogs that was measured in our experiments \cite{Aihara_2011} (see Sec. \ref{sec:method_phase}).
%A call number is calculated as the total number of calls that were included in the four-hours audio data of respective frogs.
An inter-call interval is calculated as $\delta T_{n, k} = T_{n, k+1} - T_{n, k}$ using the sequences of the call timing $T_{n, k}$
only when the condition $T_{n, k+1} - T_{n, k} \le 0.6$ sec is satisfied. 
The leader-follower relationship is determined according to the following definition: 
the leader, the $1$st follower, and the $2$nd follower are defined as the frogs that start calling first, second, and third
within the same chorus, respectively.
Note that male Japanese tree frogs start calling with low-intensity sound, 
making it difficult for us to automatically determine the leader-follower relationship.
Hence, we manually determined the leader-follower relationship by carefully looking at all the separated audio data 
and calculated the probability for taking a chorus reader for each frog.

Next, we analyze the relationship between the magnitude of attention and the three behavioral parameters 
by using a generalized linear mixed model (GLMM).
GLMM is a well-known statistical model that is used in various research areas 
so as to analyze the effects of multiple explanatory variables on a response variable \cite{Schall_1991, Faraway_2006}.
Here, we treat the magnitude of the attention paid from the $n$th frog to the $m$th frog (see Eq.(\ref{eq:kullback_leibler})) as a response variable $Y_{\mathrm{att}}$.
We then treat the three behavioral parameters 
(i.e., the inter-frog distance between the $n$th and $m$th frogs, the inter-call interval of the $m$th frog, and the leader probability of the $m$th frog) 
as explanatory variables of fixed factors, 
and describe them as $X_{\mathrm{dis}}$, $X_{\mathrm{int}}$ and $X_{\mathrm{prob}}$, respectively.
In addition, we treat frog index and experimental date as explanatory variables of random factors 
($\xi_{\mathrm{frog}}$ and $\xi_{\mathrm{date}}$) 
because these factors are difficult to be quantified but very likely to affect the calling behavior of male frogs. 
Here, we assume that the random factors $\xi_{\mathrm{frog}}$ and $\xi_{\mathrm{date}}$ follow 
a normal distribution of zero mean value 
with standard deviations of $\sigma_{\mathrm{frog}}$ and $\sigma_{\mathrm{date}}$, respectively.
By using these variables, we construct the following GLMM:
\begin{eqnarray}
	\log{\alpha} = \beta_{0} + \beta_{\mathrm{dis}}X_{\mathrm{dis}} + \beta_{\mathrm{int}} X_{\mathrm{int}} + \beta_{\mathrm{prob}} X_{\mathrm{prob}} + \xi_{\mathrm{frog}} + \xi_{\mathrm{date}},
	\label{eq:link}\\
	Y_{\mathrm{att}} \sim \mathrm{Gamma}(\alpha \beta, \beta).
	\label{eq:dist}
\end{eqnarray}
Here, the response variable $Y_{\mathrm{att}}$ is always positive
because of its definition (see Eq.(\ref{eq:kullback_leibler})).
To reproduce this feature, we assume that (1) the response variable follows a gamma distribution that always takes a positive value,
and (2) the parameter $\alpha$ of the gamma distribution
(this parameter gives the mean value of the distribution) 
is linked to the explanatory variables with a log function;  
this framework is consistent with a traditional gamma regression with multiple explanatory variables.
We confirmed that there is no multicollinearity among any pairs of the explanatory variables of fixed factors
(the absolute value of Pearson's correlation coefficient is less than $0.36$).
Posterior distributions of all the unknown parameters (i.e., $\beta_{0}$, $\beta_{\mathrm{dis}}$, $\beta_{\mathrm{int}}$, $\beta_{\mathrm{prob}}$, $\beta$, $\sigma_{\mathrm{date}}$, and $\sigma_{\mathrm{site}}$)  
were estimated from MCMC samples generated by R Statistical Software (ver. 3.4.2) and Stan (ver. 2.17.2).
Note that we normalized the explanatory variables of fixed factors from $0$ to $1$
prior to the calculation of the MCMC samples, 
%we generated 4 sets of \MOD{5000} samples, and used latter \MOD{2500} samples to calculate the posterior distributions. 
and confirmed the convergence of the MCMC samples by using $\hat{R}$ with a threshold of $1.01$ \cite{Gelman_2003}. 
The posterior mean and the 95$\%$ confidence interval of $\beta_{0}$, $\beta_{\mathrm{dis}}$, $\beta_{\mathrm{int}}$, and $\beta_{\mathrm{prob}}$
are shown in Figure~\ref{table:1}.

\section*{Acknowledgement}
We thank K. Aihara, H.G. Okuno, R. Takeda, T. Mizumoto, H. Awano, K. Itoyama, and Y. Bando 
for their valuable comments on this study.

%\section*{Data Accessibility}

%In this study, we utilize the empirical data obtained from our previous study \cite{Aihara_2011}.
%The time series data of call timing is available from the Supplementary Information of this manuscript.

\section*{Author Contributions}
K.O., I.A. and T.A. designed the research;
K.O. and I.A. analyzed the data;
K.O. performed simulation; 
K.O., I.A. and T.A. wrote the paper.

\section*{Competing Interests}
We declare we have no competing interests.

\section*{Funding}

This study was partially supported by JSPS Grant-in-Aid for Challenging Exploratory Research (No. 16K12396) 
and Grant-in-Aid for Young Scientists (No. 18K18005) to I.A.

\bibliography{MS}

\begin{thebibliography}{10}
\providecommand{\url}[1]{\texttt{#1}}
\providecommand{\urlprefix}{URL }
\expandafter\ifx\csname urlstyle\endcsname\relax
  \providecommand{\doi}[1]{doi:\discretionary{}{}{}#1}\else
  \providecommand{\doi}{doi:\discretionary{}{}{}\begingroup
  \urlstyle{rm}\Url}\fi
\providecommand{\bibAnnoteFile}[1]{%
  \IfFileExists{#1}{\begin{quotation}\noindent\textsc{Key:} #1\\
  \textsc{Annotation:}\ \input{#1}\end{quotation}}{}}
\providecommand{\bibAnnote}[2]{%
  \begin{quotation}\noindent\textsc{Key:} #1\\
  \textsc{Annotation:}\ #2\end{quotation}}
\providecommand{\eprint}[2][]{\url{#2}}

\bibitem{Vicsek_2012}
Vicsek T, Zafeiris A (2012) Collective motion.
\newblock Physics Reports 517: 71-140.
\bibAnnoteFile{Vicsek_2012}

\bibitem{Simmons_2002}
Simmons A, Fay R (2002) Acoustic communication.
\newblock Berlin: Springer Science and Business Media.
\bibAnnoteFile{Simmons_2002}

\bibitem{Bird_2003}
Catchpole C, Slater P (2003) Bird song: biological themes and variations.
\newblock Cambridge: Cambridge university press.
\bibAnnoteFile{Bird_2003}

\bibitem{Gerhardt_2002}
Gerhardt CH, Huber F (2002) Acoustic communication in insects and anurans.
\newblock Chicago: University of Chicago Press.
\bibAnnoteFile{Gerhardt_2002}

\bibitem{Wells_2007}
Wells KD (2007) The Ecology and Behavior of Amphibians.
\newblock Chicago: The University of Chicago Press.
\bibAnnoteFile{Wells_2007}

\bibitem{Takahashi_2013}
Takahashi DY, Narayanan DZ, Ghazanfar AA (2013) Coupled oscillator dynamics of
  vocal turn-taking in monkeys.
\newblock Current Biology 23: 2162-2168.
\bibAnnoteFile{Takahashi_2013}

\bibitem{Demartsev_2018}
Demartsev V, Strandburg-Peshkin A, Ruffner M, Manser M (2018) Vocal turn-taking
  in meerkat group calling sessions.
\newblock Current Biology 28: 3661-3666.
\bibAnnoteFile{Demartsev_2018}

\bibitem{Bronkhorst_2000}
Bronkhorst AW (2000) The cocktail party phenomenon: A review of research on
  speech intelligibility in multiple-talker conditions.
\newblock Acta Acustica united with Acustica 86: 117-128.
\bibAnnoteFile{Bronkhorst_2000}

\bibitem{Fish_2011}
Katza Y, Tunstroma K, Ioannoua CC, Huepeb C, Couzin ID (2011) Inferring the
  structure and dynamics of interactions in schooling fish.
\newblock PNAS 108: 18720-18725.
\bibAnnoteFile{Fish_2011}

\bibitem{Bird_2008}
Ballerini M, Cabibbo N, Candelier R, Cavagna A, Cisbani E, et~al. (2008)
  Interaction ruling animal collective behavior depends on topological rather
  than metric distance: Evidence from a field study.
\newblock PNAS 105: 1232-1237.
\bibAnnoteFile{Bird_2008}

\bibitem{Fujioka_2016}
Fujioka E, Aihara I, Sumiya M, Aihara K, Hiryu S (2016) Echolocating bats use
  future target information for optimal foraging.
\newblock PNAS 113: 4848-4852.
\bibAnnoteFile{Fujioka_2016}

\bibitem{Maeda_1999}
Maeda N, Matsui M (1999) Frogs and Toads of Japan.
\newblock Tokyo: Bunichi Sogo Shuppan Co. Ltd.
\bibAnnoteFile{Maeda_1999}

\bibitem{Aihara_2009}
Aihara I (2009) Modeling synchronized calling behavior of japanese tree frogs.
\newblock Physical Review E 80: 011918.
\bibAnnoteFile{Aihara_2009}

\bibitem{Aihara_2011}
Aihara I, Takeda R, Mizumoto T, Otsuka T, Takahashi T, et~al. (2011) Complex
  and transitive synchronization in a frustrated system of calling frogs.
\newblock Physical Review E 83: 031913.
\bibAnnoteFile{Aihara_2011}

\bibitem{Aihara_2014}
Aihara I, Mizumoto T, Otsuka T, Awano H, Nagira K, et~al. (2014)
  Spatio-temporal dynamics in collective frog choruses examined by mathematical
  modeling and field observations.
\newblock Scientific Reports 4: 3891.
\bibAnnoteFile{Aihara_2014}

\bibitem{Aihara_2016}
Aihara I, Mizumoto T, Awano H, Okuno HG (2016) Call alternation between
  specific pairs of male frogs revealed by a sound-imaging method in their
  natural habitat.
\newblock Proceedings of Interspeech 2016 : 2597-2601.
\bibAnnoteFile{Aihara_2016}

\bibitem{Bee_2013}
Bee MA, Schwartz JJ, Summers K (2013) All's well that begins wells: celebrating
  60 years of animal behaviour and 36 years of research on anuran social
  behaviour'.
\newblock Animal Behaviour 85: 5-18.
\bibAnnoteFile{Bee_2013}

\bibitem{Kuramoto_1984}
Kuramoto Y (1984) Chemical Oscillations, Waves, and Turbulence.
\newblock Berlin: Springer-Verlag.
\bibAnnoteFile{Kuramoto_1984}

\bibitem{RSOS_2018}
Aihara I, Kominami D, Hirano Y, Murata M (2019) Mathematical modelling and
  application of frog choruses as an autonomous distributed communication
  system.
\newblock Royal Society Open Science 6: 181117.
\bibAnnoteFile{RSOS_2018}

\bibitem{Strogatz}
Strogatz S (2014) Nonlinear Dynamics and Chaos (2nd Edition).
\newblock Boca Raton: CRC Press.
\bibAnnoteFile{Strogatz}

\bibitem{Nenkin}
Takamatsu A, Fujii T, Endo I (2000) Time delay effect in a living coupled
  oscillator system with the plasmodium of {P}hysarum polycephalum.
\newblock Physical Review Letters 85: 2026.
\bibAnnoteFile{Nenkin}

\bibitem{BZ}
Miyazaki J, Kinoshita S (2006) Determination of a coupling function in
  multicoupled oscillators.
\newblock Physical Review Letters 96: 194101.
\bibAnnoteFile{BZ}

\bibitem{Walk}
Funato T, Yamamoto Y, Aoi S, Imai T, Aoyagi T, et~al. (2016) Evaluation of the
  phase-dependent rhythm control of human walking using phase response curves.
\newblock PLoS Computational Biology 12 (5): e1004950.
\bibAnnoteFile{Walk}

\bibitem{Ota_arXiv}
Ota K, Aoyagi T (2014) Direct extraction of phase dynamics from fluctuating
  rhythmic data based on a bayesian approach : arXiv:1405.4126.
\bibAnnoteFile{Ota_arXiv}

\bibitem{Brush_1989}
Brush JS, Narins PM (1989) Chorus dynamics of a neotropical amphibian
  assemblage: comparison of computer simulation and natural behaviour.
\newblock Animal Behaviour 37-1: 33-44.
\bibAnnoteFile{Brush_1989}

\bibitem{Jones_2014}
Jones DL, Jones RL, Ratnam R (2014) Calling dynamics and call synchronization
  in a local group of unison bout callers.
\newblock Journal of Comparative Physiology A 200-1: 93-107.
\bibAnnoteFile{Jones_2014}

\bibitem{Grafe_1996}
Grafe TU (1996) The function of call alternation in the {A}frican reed frog
  ({H}yperolius marmoratus): precise call timing prevents auditory masking.
\newblock Behavioral Ecology and Sociobiology 38: 149-158.
\bibAnnoteFile{Grafe_1996}

\bibitem{Snedden_1998}
Snedden WA, Greenfield MD, Jang Y (1998) Mechanisms of selective attention in
  grasshopper choruses: who listens to whom?
\newblock Behavioral Ecology and Sociobiology 43: 59-66.
\bibAnnoteFile{Snedden_1998}

\bibitem{CallRate_1992}
Ryan MJ, Keddy-Hector A (1992) Directional patterns of female mate choice and
  the role of sensory biases.
\newblock The American Naturalist 139: S4-S35.
\bibAnnoteFile{CallRate_1992}

\bibitem{Buck_1968}
Buck J, Buck E (1968) Mechanism of rhythmic synchronous flashing of fireflies.
\newblock Science 159: 1319.
\bibAnnoteFile{Buck_1968}

\bibitem{Strogatz_2004}
Strogatz S (2004) Sync: the emerging science of spontaneous order.
\newblock London: Penguin.
\bibAnnoteFile{Strogatz_2004}

\bibitem{Backwell_1998}
Backwell P, Jennions M, Passmore N, Christy J (1998) Synchronized courtship in
  fiddler crabs.
\newblock Nature 391: 31-32.
\bibAnnoteFile{Backwell_1998}

\bibitem{Backwell_1999}
Backwell P, Jennions M, Christy J, Passmore N (1999) Female choice in the
  synchronously waving fiddler crab {U}ca annulipes.
\newblock Ethology 105: 415-421.
\bibAnnoteFile{Backwell_1999}

\bibitem{Markham_2009}
Markham M, McAnelly M, Stoddard P, Zakon H (2009) Circadian and social cues
  regulate ion channel trafficking.
\newblock PLoS Biology 7: e1000203.
\bibAnnoteFile{Markham_2009}

\bibitem{Piko}
Pikovsky A, Rosenblum M, Kurths J (Cambridge) Synchronization: A Universal
  Concept in Nonlinear Sciences.
\newblock 2001: Cambridge University Press.
\bibAnnoteFile{Piko}

\bibitem{Schall_1991}
Schall R (1991) Estimation in generalized linear-models with random effects.
\newblock Biometrika 78: 719-727.
\bibAnnoteFile{Schall_1991}

\bibitem{Faraway_2006}
Faraway J (2006) Extending the Linear Model with R.
\newblock Boca Raton: CRC Press.
\bibAnnoteFile{Faraway_2006}

\bibitem{Gelman_2003}
Gelman A, Carlin J, Stern H, Rubin D (2003) Bayesian Data Analysis.
\newblock Boca Raton: Chapman and Hall/CRC.
\bibAnnoteFile{Gelman_2003}

\end{thebibliography}

\clearpage

\clearpage

\end{document}